\documentclass[12pt]{article}  

\usepackage{amsmath,amsfonts,amssymb}
\usepackage{graphicx}
\usepackage{float}
\usepackage[colorlinks=true, allcolors=blue]{hyperref}

\providecommand{\keywords}[1]
{
  \small	
  \textbf{\textit{Keywords---}} #1
}

\title{Spectropolarimeter's optical design for the Arago space mission project}

\author{
{Eduard Muslimov}$^{a,b}$,{Coralie Neiner}$^{c}$\\
\small a--{NOVA Optical IR Instrumentation Group at ASTRON,}\\
\small {Oude Hoogeveensedijk 4, 7991 PD Dwingeloo, The Netherlands}\\
\small b--{Aix Marseille Univ, CNRS, CNES, LAM, Marseille, France}\\
\small c--{LESIA, Observatoire de Paris, 5 pl. Jules Janssen, 92195 Meudon, France}\\
}

\begin{document} 
\maketitle{}

\begin{abstract}
    Arago is a concept of space mission submitted to the European Space Agency’s M7 science program. It will target a number of science cases in stellar physics including the characterisation of star-planet interactions. The concept is based on a 1-m class Ritchey-Chretien F/13 telescope mounted on an Ariel-type platform. The scientific payload includes a common polarimetric unit and 2 spectrographs connected via a dichroic splitter. The polarimetric unit consists of 6 $MgF_2$ plates in pairs connected by optical contact and a Wollaston analyzer. Each of the spectral channels represents an echelle spectrograph. The first one will operate in the ultraviolet range 119-320 nm with the spectral resolving power of $R\geq 25 000$. It consists of an off-axis parabolic collimator, an echelle grating in a quasi-Littrow mounting, a cross-disperser concave grating, and a CMOS camera. The cross-disperser grating works also as a camera mirror and represents a holographic grating imposed on a spherical substrate. It is recorded holographically with an auxiliary deformable mirror to correct the aberrations and has a blazed grooves profile. The spectral image is projected onto a $\delta$-doped CMOS detector. The second spectral channel operates in the visible range 350-888 nm with $R\geq 35 000$. Its design uses an immersed grating, i.e. a blazed reflective grating ruled on the backside of a fused silica prism. This solution should allow us to operate in a spectral range wider than one octave with sufficient spectral orders separation. The dispersed beams are focused with a 4-lens objective onto a CMOS detector. In addition to the main scientific payload, the optical design includes two stages of fine-guiding system. The first stage represents a simple projecting system tracking the image around the entrance pinhole and communicating with the platform actuators. The second stage is fed by the $0^{th}$ diffraction order of the visible channel echelle grating. Its information is communicated to a tip-tilt mirror in front of the dichroic. The first stage should improve the pointing accuracy from $8"$ to \textit{200 mas} precision to compensate the platform jitter and drift and guarantee that the star images passes through the instrument’s entrance pinhole. The second stage should correct the pointing accuracy further as well as some thermo-elastic deformations. 
\end{abstract}


\section{INTRODUCTION}
\label{sec:intro}  
\textit{Arago} is a concept of mid-scale space mission focused on high-resolution spectropolarimerty. The goal of \textit{Arago} is to follow the cycle of matter, and therefore the entire life of stars and planets from their formation from interstellar gas and grains to their death and feedback into the interstellar medium (ISM). During the formation and throughout the life of stars and planets, a few key basic astrophysical properties, especially magnetic fields, stellar winds, rotation, and binarity, influence their dynamics, and thus fundamentally impact their evolution.
The associated processes directly affect the internal structure of stars, the dynamics, and the immediate circumstellar environment. They consequently drive stellar evolution, but also define the environments of planets, thereby influencing the formation and fate of planets surrounding the stars. \textit{Arago} will allow us to obtain, for the first time, a full picture of the 3D dynamical environment of stars and their interactions with planets, and explore the conditions for the emergence of life on exoplanets. To reach this aim, \textit{Arago} will perform high-resolution spectropolarimetry simultaneously in the ultra-violet (UV) and visible (VIS) wavelength ranges, concentrating on two groups of science goals: study of the cycle of matter in the Milky Way and  interaction between the stars and their planets. 

To reach these science goals, it is necessary to observe stellar spectral lines and their polarisation in the two spectral domains simultaneously and with a high-cadence continuous monitoring. The Visible domain allows us to characterise the surface of the star: its properties (e.g., temperature, gravity, rotation, magnetic field) and surface features (e.g., spots, chemical enhancements). The UV domain allows us to characterise the environment of the star: its wind, magnetosphere, chromosphere, irradiation of exoplanets, etc. Observing both domains simultaneously is the only way to obtain a complete 3D view of the star and its surroundings, and directly link surface features to circumstellar structures, e.g. surface spots to coronal mass ejections, or magnetic footpoints to discs. In addition, the measurement of polarisation in the spectra allows us to detect and quantify the magnetic field and environment of stars and exoplanets. Linear polarimetry provides a means to determine deviations from spherical symmetry of all kinds of objects. It is the extension of interferometry into the domain that is not restricted by the objects’ angular size but merely by its flux \cite{udDoula2022, Morin2019}.

The optical design of the mission has to match a number of the contradictory requirements. On the one hand it should provide high spectral resolution in a wide spectral range and have a high throughput, especially in the UV band to reach a good signal-to-noise ration (SNR). In addition, it should minimize the impact of the spectral image combination, as well as instabilities of the instrument and spacecraft. On the other hand, being a mid-scale mission, \textit{Arago} must rely on existing technologies and elements to reduce technological risks and make it possible to built and launch the mission in the coming decade. Also, as any space mission payload it should be as simple, stable, compact and lightweight, as possible.

Below we present the optical architecture of  \textit{Arago}, which is driven by the science goals and the corresponding technical specifications \cite{Pertenais2017}. Then we describe design of every subsystem, which meets the above-stated requirements to the  performance and technological readiness. 

\section{OPTICAL SYSTEM ARCHITECTURE}
\label{sec:ARCH}  
Taking into account the science goals and the available spacecraft platform we can figure out the following top-level requirements for the optical design:
\begin{itemize}
    \item The mission should be based on a 1-m, F/13 telescope.
    \item The polarimeter should be working for the wide working spectral range of 123-888 nm from the UV to VIS without exchange or rectaction.
    \item The spectral part should consist of 2 branches - UV (119-320 nm, $R\geq 25 000$) and VIS (355-888 nm, $R\geq 35 000$).
    \item The payload should fit behind the structure of the telescope primary mirror.
    \item The spectral image should be stable with an accuracy of 1 $\mu m$ at the detector during $\leq 30$ min acquisition time.
\end{itemize}

\begin{figure} [ht]
   \begin{center}
   \begin{tabular}{c} 
   \includegraphics[width=12cm]{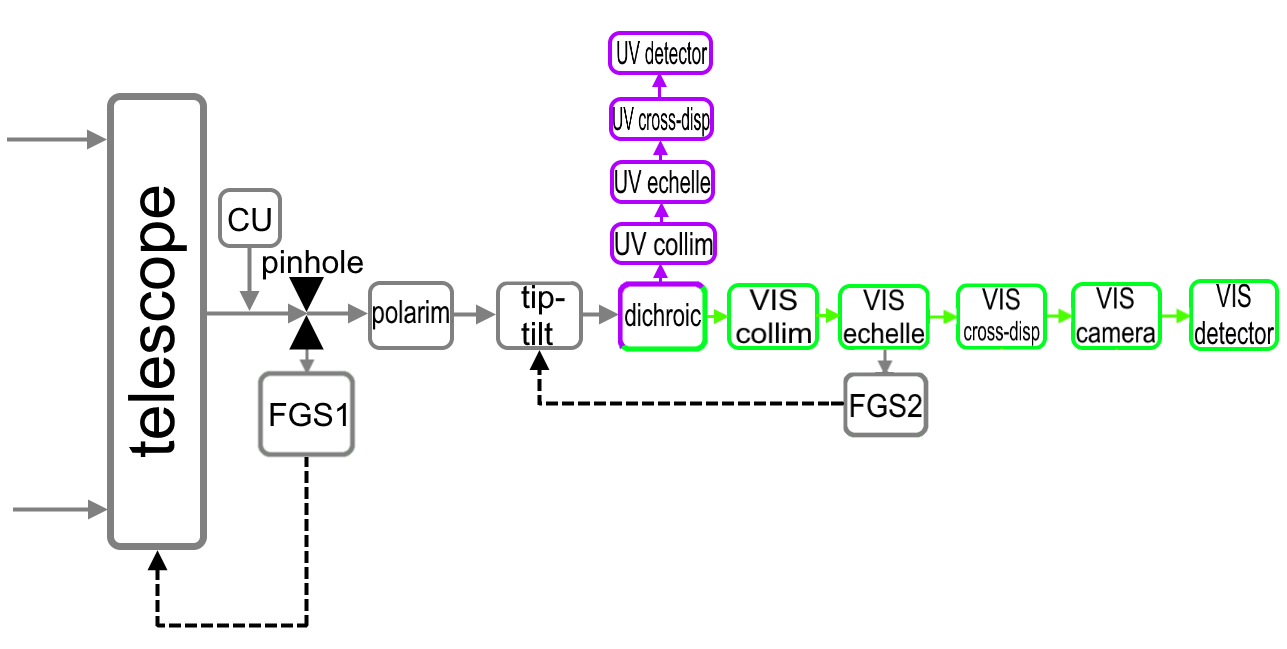}
   \end{tabular}
   \end{center}
   \caption[example] 
   { \label{fig:arch} 
Optical architecture of \textit{Arago} space mission scientific payload.}
   \end{figure}

The top-level architecture of the optical system is shown in Fig.~\ref{fig:arch}. The 1-m diameter on-axis Cassegrain telescope is oriented towards a point-source science target and images the star on an entrance pinhole. The telescope uses a conventional optical design, which is not described here in details. The distance between the two mirrors is 1209.3 mm, the central obscuration is $1.27\%$ and the maximum field of view is $\pm 0.26'$. 

The platform provides coarse pointing with an error of $8"$. We use a fine guiding system (FGS) to compensate it. The first stage of the FGS records a linear field of view of 2 mm in diameter (i.e. $32"$) on the mirror around the entrance pinhole. An attitude and orbital control system (AOCS) loop between the FGS and platform allows us to refine the platform pointing to 200 mas, placing and stabilizing the target image on the hole with a precision of $12 \mu m$. Taking into account the Airy disk diameter in the diffraction-limited point spread function (PSF) formed by the telescope, which reaches $28 \mu m$, we obtain the  diameter of the pinhole equal to $53\mu m$. The hole isolates the target from the surroundings and guarantees an upper limit for the image size. 

The F/13 beam from the telescope then enters the polarimeter, placed just after this field stop. The polarimeter consists of a modulator followed by a polarisation separator. At the exit of the polarimeter the beam is bent by a tip-tilt mirror, which is used as the actuator of the FGS second stage. Also  it allows to implement a packaging scheme, where the payload is mounted on a single baseplate parallel to the telescope primary mirror backside. Then the beam is split by a dichroic plate, which reflects the UV light in the wavelength range 119-320 nm and transmits the VIS wavelength range 355-888 nm. 

The two beams then enter their respective spectrograph and detection chains. Each of the spectral channels represents an echelle spectrograph, consisting of reflective collimator, echelle grating as the main disperser, cross-disperser, camera, and detector. In order to minimize the number of reflection in the UV channel we consider merging the functions of cross-disperser and camera. The VIS and UV detection chains are similar, making use of a CMOS sensor, but optimized with the appropriate size, pixel pitch, and a $\delta$-doping for the UV channel. 

The second stage of the FGS records the $0^{th}$ order of the VIS spectrum on a small CMOS sensor. A tip-tilt mirror placed between the polarimeter and the dichroic uses this information to maintain the image at a stable position on the FGS detector, hence also maintaining the science orders position fixed on their detectors. This second FGS stage allows us to correct for remaining pointing errors as well as thermo-elastic effects in most of the instrument, thus guaranteeing the stability of the science spectrum on the detectors at the required sub-pixel level.

Finally, the design also contains a calibration unit (CU), which can be coupled with the scientific payload with a retractable mirror  mounted right before the entrance pinhole. However, the CU is not used for the main operation modes, so it is not considered below.
In the following sections we consider the optical design and nominal performance of the listed sub-systems one-by-one.

\section{SPECTROGRAPHS}
\label{sec:SPEC}  

\subsection{Spectrographs design}
\label{sec:spe_design}

\begin{figure} [ht]
   \begin{center}
   \begin{tabular}{c} 
   \includegraphics[width=12cm]{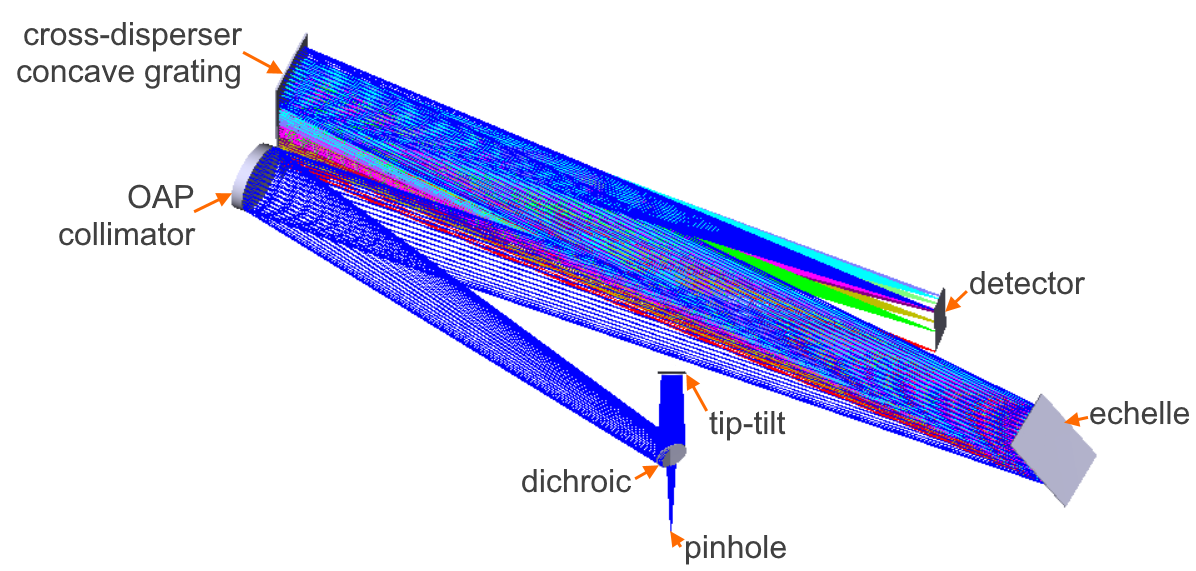}
   \end{tabular}
   \end{center}
   \caption[example] 
   { \label{fig:uv} 
General view of the UV spectrograph optical design.}
   \end{figure}

The two spectrographs designs are similar. The UV spectrograph design is shown in Fig.~\ref{fig:uv}. It grants simultaneous spectroscopy in the spectral range from 119 to 320 nm. 

The spectrum reflected from the dichroic goes to an off-axis parabolic (OAP) collimator that sends the beam to an echelle grating. The main disperser works in 25 diffraction orders (from 15 to 39). The grating frequency and its blazing angle are reachable with the currently available technologies\cite{Kruczek22}. Note, that  the free dispersion range in the shortest order is larger than 3 nm. This will avoid having too many spectral lines cut over two consecutive orders and increase the accuracy of spectropolarimetric data, so the spectral order length was one of the important requirements for the spectrographs design. 

\begin{figure} [ht]
   \begin{center}
   \begin{tabular}{c} 
   \includegraphics[width=12cm]{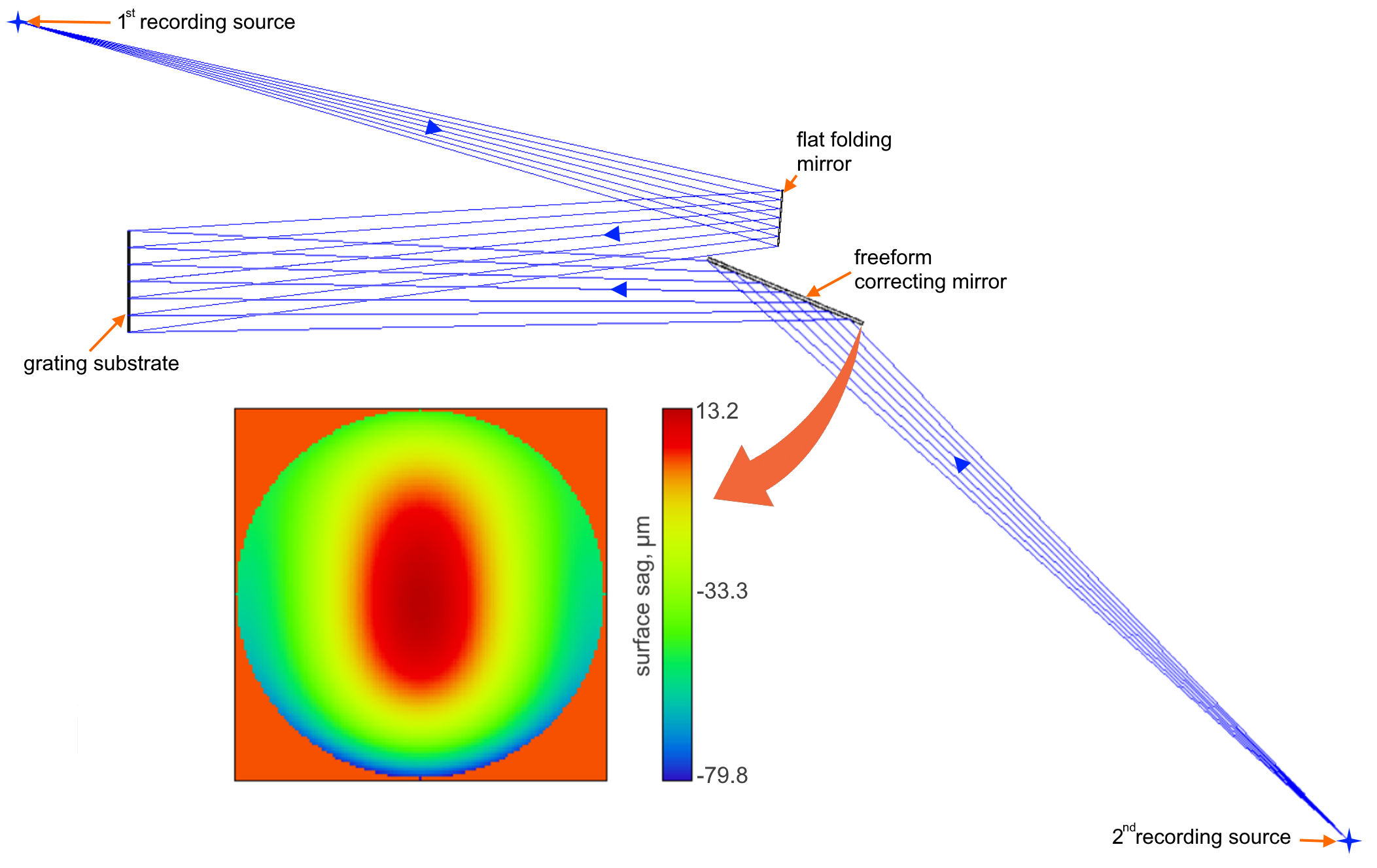}
   \end{tabular}
   \end{center}
   \caption[example] 
   { \label{fig:rec} 
Design of the recording setup for the UV cross-disperser using an auxiliary mirror.}
   \end{figure}

The cross-dispersion and focusing of the beam is made with a spherical concave grating. With the given spectral range and target dispersion and resolution it is necessary to introduce additional degrees of freedom to compensate the aberrations of the concave grating, which is mounted in a dispersed beam \cite{Gatto2017}. Thus, the grating is recorded with one spherical and one aberrated wavefront at 488 nm. The said aberrated wavefront can be formed by a deformable mirror as it is shown in Fig.~\ref{fig:rec}. In this example the recording point sources are placed 1100 mm away from the substrate and the auxiliary mirror is mounted in the second beam with the angle of incidence (AOI) of $68.4^\circ$. The surface shape is described by Zernike polynomials\cite{Lakshminarayanan2011} $Z_4,Z_6,Z_7,Z_9,Z_{11}$ and $Z_{12}$ with the peak-to-valley (PTV) surface sag of $93.1 \mu m$. The grooves profile obtained in such a recording setup can be  turned into a blazed triangular shape either by recording in oppositely propagating beams \cite{Gatto2017} (e.g. when the first recording beam is converging and the second one is diverging) or by ion-etching \cite{Znamenskiy07}. The linear magnification factor between the collimator and camera is \textit{0.8}. 

The full spectral range is fitted within a single image on a CMOS detector that fulfils the requirements in terms of optical quality for the proper separation of the two polarised beams. This design is tailored to fit on a detector with $4608 \times 1920$ pixels and $16 \mu m $ pixel pitch \cite{Bai08, Wang2020} , thus providing the pinhole image sampling of 2.65, which is above the Shannon-Nyquist criterion \cite{Shannon1949}.

\begin{figure} [ht]
   \begin{center}
   \begin{tabular}{c} 
   \includegraphics[width=12cm]{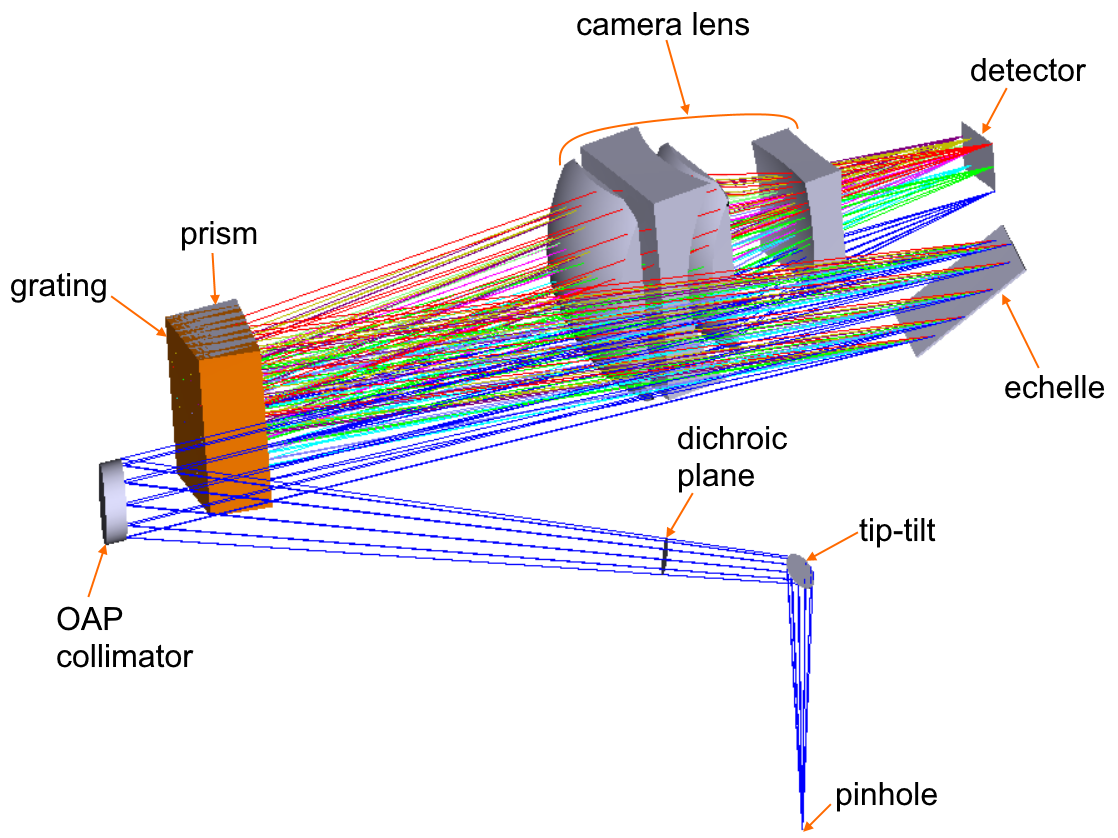}
   \end{tabular}
   \end{center}
   \caption[example] 
   { \label{fig:vis} 
General view of the VIS spectrograph optical design.}
   \end{figure}

Similarly, the VIS spectrograph design is shown in Fig.~\ref{fig:vis}.
It works in a wide spectral range from 355 to 888 nm. Following the
dichroic plate and an OAP collimator, the collimated beam is spectrally dispersed by the echelle grating to  37  orders  (24 to 60). The minimum spectral length of the order is 5.8 nm in this case. To separate this number of orders and limit the size and spatial frequency of the cross-disperser we propose to use an immersed grating. It represents a blazed reflective grating imposed on rear side of a fused silica prism \cite{Heusinger22, Amerongen2010}. The spectral image is focused by a Tessar-type camera with 4 lenses. The lenses are uncemented, the front surface of the $1^{st}$ one is a $2^{nd}$ order asphere with the PTV deviation from the best fit sphere of $12.8 \mu m$, and the back side of the $4^{th}$ lens is a $6^{th}$ order asphere with the PTV asphericity of $60.6 \mu m$. Note that due to the spectral image format the clear aperture of the lenses is close to a rectangle. The  linear magnification factor between the collimator and the camera is \textit{0.491}. The spectral image is focused onto a CMOS detector with $4096 \times 4096$ pixel and $10 \mu m$ pixel pitch. The sampling of the pinhole image is slightly different, but still satisfies the  Shannon-Nyquist criterion.

Finally, all the design parameters for the two spectrographs are summarized in Table~\ref{tab:param}. All of the parameters are achievable with the currently existing devices and technologies.

\begin{table}[!ht]
\caption{Design parameters of the \textit{Arago} spectrographs} 
\label{tab:param}
\begin{center}       
\begin{tabular}{|l|c|c|} 
\hline
  \textbf{Channel} & \textbf{UV} & \textbf{VIS} \\
\hline
  Wavelength range, nm & 119-320 &	350-888 \\
\hline
 Min. spectral resolving power	& 25 000	& 35 000\\
\hline
  \multicolumn{3}{|c|}{\textbf{Collimator}}\\
\hline
  Focal length, mm		& 850 & 700\\
\hline
  Clear aperture, mm	& 68 & 59\\
\hline
  Decenter, mm	& 245.1	& 375.1\\
\hline
  Beam tilt, $^\circ$ & 17	& 30\\
\hline
  \multicolumn{3}{|c|}{\textbf{Echelle}}\\
\hline
  Frequency, $mm^{-1}$ & 281	& 79\\
\hline
  AOI, $^\circ$ & 40.61	& 55.55\\
\hline
  Clear aperture X$\times$ Y, mm	& 68 $\times$ 89	& 59 $\times$ 103\\
\hline
  Orders & 15-39	& 24-60\\
\hline
  Free order length, nm & 3.06-20.70 & 5.80-36.27\\
\hline
  Full order length, nm & 7.96-20.70 & 14.51-36.27\\
\hline
  \multicolumn{3}{|c|}{\textbf{Cross-disperser}}\\
\hline
  Type	 & Concave grating  & Immersed grating\\
\hline
  Clear aperture X $\times$ Y, mm		& 160 $\times$ 72	  & 132 $\times$ 70\\
\hline
  Substrate		& Spherical mirror	  & F.silica $7.58^\circ$ prism\\
\hline
  Frequency, $mm^{-1}$ & 232.45	& 208 \\
\hline
  \multicolumn{3}{|c|}{\textbf{Camera}}\\
\hline
  Type & Concave grating & 4-lens objective \\
\hline
  Focal length, mm & 671.9 & 343.4 \\
\hline
  Clear aperture X $\times$ Y, mm	 & 160 $\times$ 72 & 166 $\times$  104 \\
\hline
  \multicolumn{3}{|c|}{\textbf{Detector}}\\
\hline
  Type	 & d-doped CMOS	 & CMOS \\
\hline
  Size, mm	 & 73.73 $\times$ 30.7	 & 40.9 $\times$ 40.9 \\
\hline
  Pixel size, $\mu m$		 & 16	 & 10 \\
\hline
  Sampling & 2.65 & 2.6\\
\hline
\end{tabular}
\end{center}
\end{table} 

\subsection{Spectrographs performance}
\label{sec:spe_perf}

\begin{figure} [ht]
   \begin{center}
   \begin{tabular}{c} 
   \includegraphics[width=13cm]{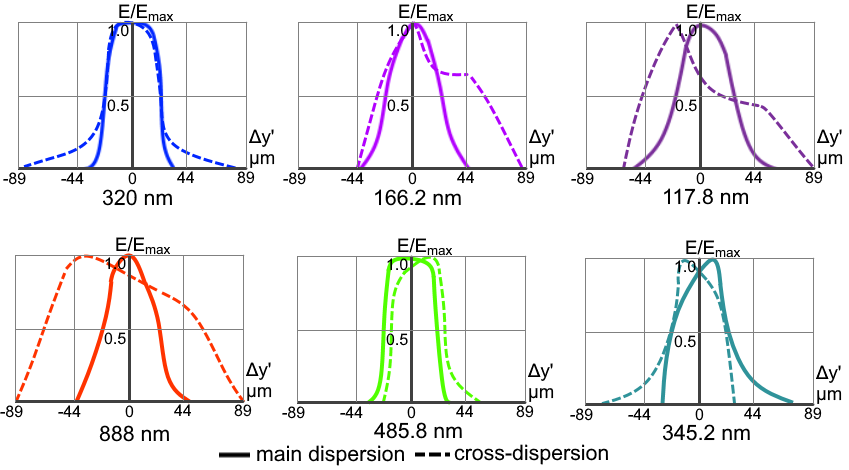}
   \end{tabular}
   \end{center}
   \caption[example] 
   { \label{fig:if} 
Sample instrument functions for the the spectrographs in 2 channels: top -- UV channel (pinhole diameter is 42.4 $\mu m$ in projection), bottom -- VIS channel (pinhole diameter is 26 $\mu m$ in projection).}
   \end{figure}

The performance of designed spectrographs is assessed mainly  through their spatial resolution metrics. The aberrations in the main dispersion direction will have a direct impact on the spectral resolution and, therefore, on the scientific performance. In the meantime, the aberrations in the cross-dispersion direction may change slightly the illumination in the pinhole image. Moreover, the images corresponding to different orders and polarization states should be separated by at least 2 dark pixels in the cross-dispersion direction. This allows to sacrifice the image quality in this direction to a certain extend and assign a 1/10 weight coefficient to the corresponding aberrations during the design optimization. The key metrics computed to analyze the spectral resolution are the spectrograph's instrument functions showing the illumination distribution in the pinhole image. The instrument functions in both of the sections for the center and the corners of the 2D spectral image of each channel are shown  in Fig.~\ref{fig:if}.

\begin{figure} [ht]
   \begin{center}
   \begin{tabular}{c} 
   \includegraphics[width=13cm]{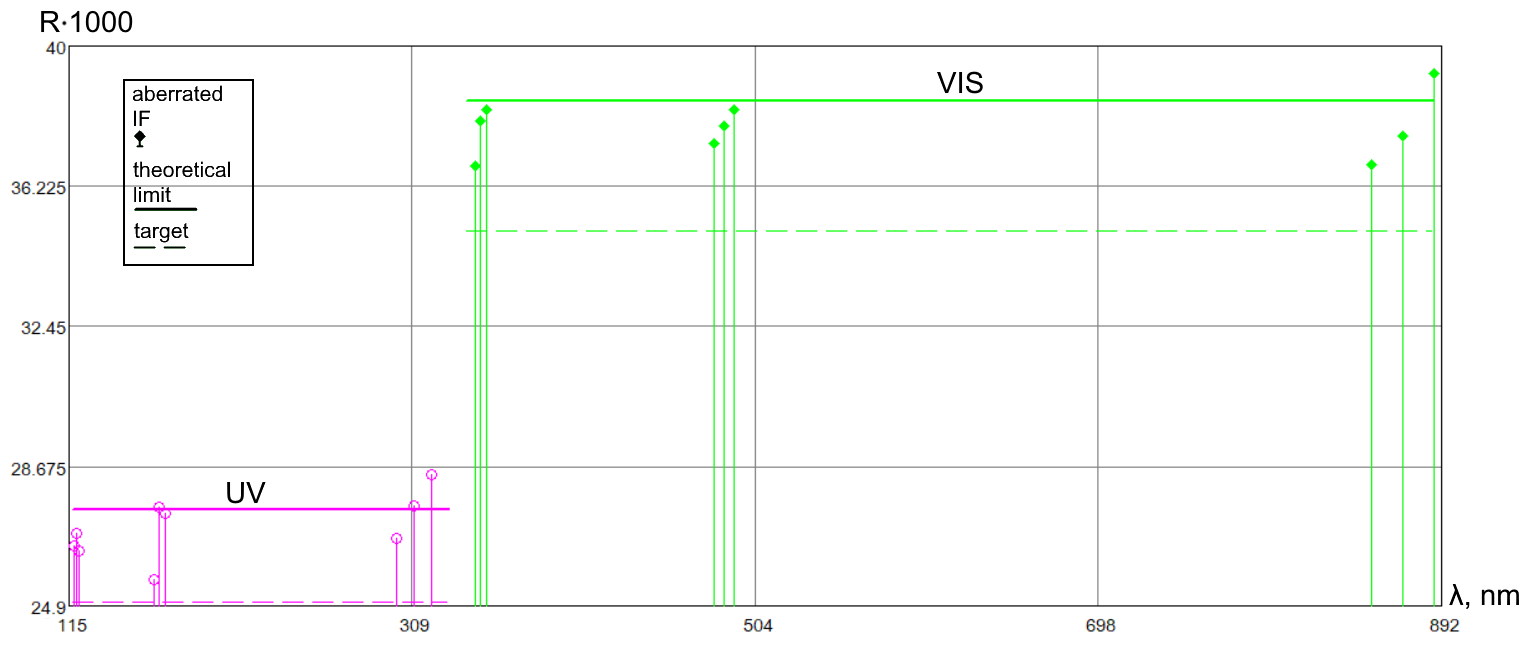}
   \end{tabular}
   \end{center}
   \caption[example] 
   { \label{fig:res} 
Spectral resolving power of the 2 channels.}
   \end{figure}

We compute the spectral resolution limit  $\delta \lambda$ as a product of the instrument function full width at half maximum (FWHM) by the reciprocal linear dispersion provided by the echelle. Then the spectral resolving power is computed as $R=\lambda/\delta \lambda$. The resulting values are shown on a diagram (Fig.~\ref{fig:res}) for 3 wavelengths in the first, middle, and last orders of each channel. As the plot shows, the spectral resolving power varies from 25 620 to 28 450 in the UV channel and from 36 770 to 39 270 in the VIS. So, the spectral resolution is above the requirements and leaves a certain margin for the manufacturing and alignment errors. However, we must note here that in both of the cases the image quality is driven by the camera part aberrations, rather than by the theoretical limit defined by the collimated beam size and the echelle parameters. Also, the design may require slight changes to compensate the residual dispersion introduced by the polarimeter.

Another key aspect of the optical system performance is its end-to-end throughput. The exact values will depend strongly on the diffractive elements groove profiles and the available coatings, but for the approximate estimation we have used a few conservative assumptions: 
\begin{itemize}
    \item The $Al+MgF_2$ telescope coating reflectance varies between $87\%$ and $95\%$ per mirror depending on wavelength \cite{Greer2012};
    \item $100\%$ of the light goes through the pinhole;
    \item The polarimeter transmission varies between $20$ and $60\%$ depending on wavelength\cite{Pertenais2015,Pertenais2016};
    \item The dichroic has $70$ to $85\%$ reflectivity in the UV and $95\%$ in the VIS\cite{Ferrari2019};
    \item The UV collimator has the same coating as the telescope while the VIS collimator has between $87\%$ and $90\%$ reflectance;
    \item The diffraction efficiencies are up to $80\%$ and $78\%$ for the UV and VIS echelles \cite{Ferrari2019}, respectively, and up to $76\%$ for both of the cross-dispersers \cite{Muslimov2018};
    \item The CMOS detectors have the quantum efficiency (QE) varying between
$30\%$ and $55\%$ in the UV and $30\%$ and $90\%$ in the VIS \cite{Bai08}.
\end{itemize}   
With all of these values we obtain the global throughput of the VIS spectrograph with its detector varying between $3$ and $33\%$ along the working range, and that for the UV channel from $7$ to $24\%$.

\section{POLARIMETER}
\label{sec:POL}  

\begin{figure} [ht]
   \begin{center}
   \begin{tabular}{c} 
   \includegraphics[width=12cm]{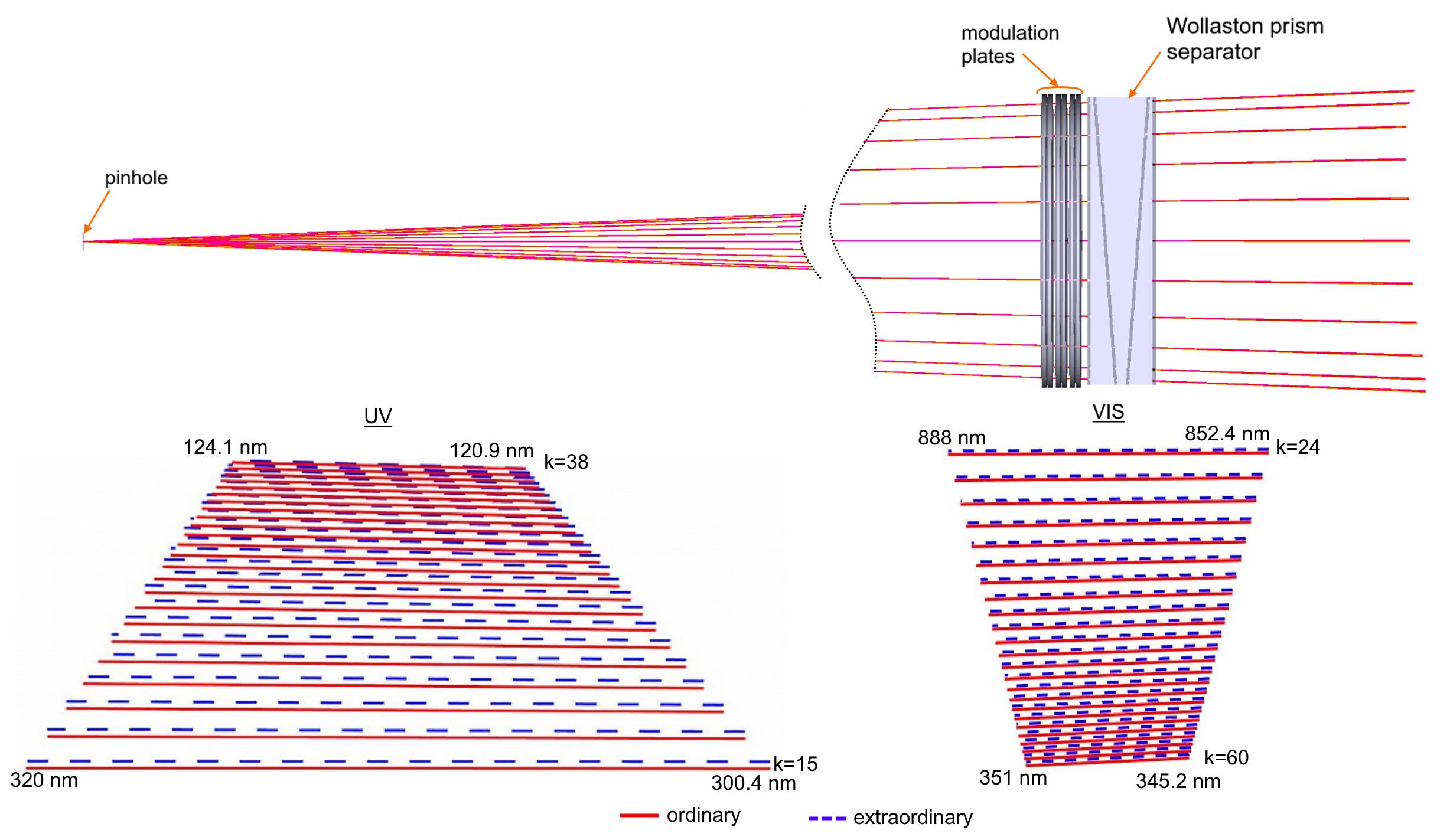}
   \end{tabular}
   \end{center}
   \caption[example] 
   { \label{fig:pol} 
Polarimeter unit design and performance: top -- optical design of the polarimeter, bottom left -- polarization components separation in the UV, bottom right -- polarization components separation in the VIS.}
   \end{figure}

The polarimeter is the key module of the instrument for \textit{Arago} and is placed directly after the Cassegrain focus of the telescope to minimise instrumental polarisation. It is composed of two different parts: the modulator and the analyser, which separates the polarization components (Fig.~\ref{fig:pol}, top).
The modulator and the analyser are made of $MgF_2$, the only material that is both birefringent and transparent in the UV. Due to the wide wavelength range, it is not possible to find polarisation components with achromatic retardation. The solution is to develop a  polarimeter with the highest efficiency of Stokes parameters extraction. The modulator is thus a stack of 3 pairs of plates, where each pair is a quasi-zero-order retarder \cite{Pertenais2015,Pertenais2016}. The 3 thickness differences and the relative angles of the plates of the modulator are adjusted to achromatise and optimise the extraction efficiencies of the Stokes parameters from 123 to 888 nm \cite{Snik2012}. This whole stack of plates rotates by steps of 30° to ensure a full-Stokes measurement. For a complete measurement, 6 sub-exposures (at 0°, 30°, 60°, 90°, 120°, and 150°) have to be used in the demodulation process. The analyser is a Wollaston prism, separating the two orthogonal polarisation states. As the birefringence of the material drops towards the FUV (and goes to 0 at 119.5 nm), we need to use a double-prism Wollaston to separate the two orthogonal beams correctly at the lowest wavelength of \textit{Arago}. Moreover, the wavelength dependence of the index of refraction of $MgF_2$ has been taken into account to size the beam separation for the two polarisation states, which is the smallest at the shortest wavelength.
This polarimeter ensures a high precision and sensitivity measurement of the full Stokes IQUV vector from 123 to 888 nm. The separation of the two polarization components in the UV varies from $170$ to $674~ \mu m$ and is relatively stable at the wavelengths after 162 nm. In the VIS channel the separation is $347-377 ~ \mu m$. In both of the cases the separation is sufficient even taking into account the astigmatic elongation of the pinhole image. The polarization components separation as well as the free dispersion zone format for the UV and VIS channels are shown in Fig.~\ref{fig:pol}, bottom.

\section{FINE GUIDING SYSTEM}
\label{sec:FGS}  

The FGS consists of two separate stages: the FGS1 insures a fine pointing stability of \textit{200 mas} and allows to bring and maintain the target light into the instrument; the FGS2 is internal to the instrument and insures the required
stability of $1 \mu m$ on the detectors during up to 30 minutes.
Using the service vehicle module (SVM) pointing system, the target is acquired with a precision of $\pm 8"$ by the star tracker.

\subsection{FGS stage 1}
\label{sec:fgs1}

\begin{figure} [ht]
   \begin{center}
   \begin{tabular}{c} 
   \includegraphics[width=13cm]{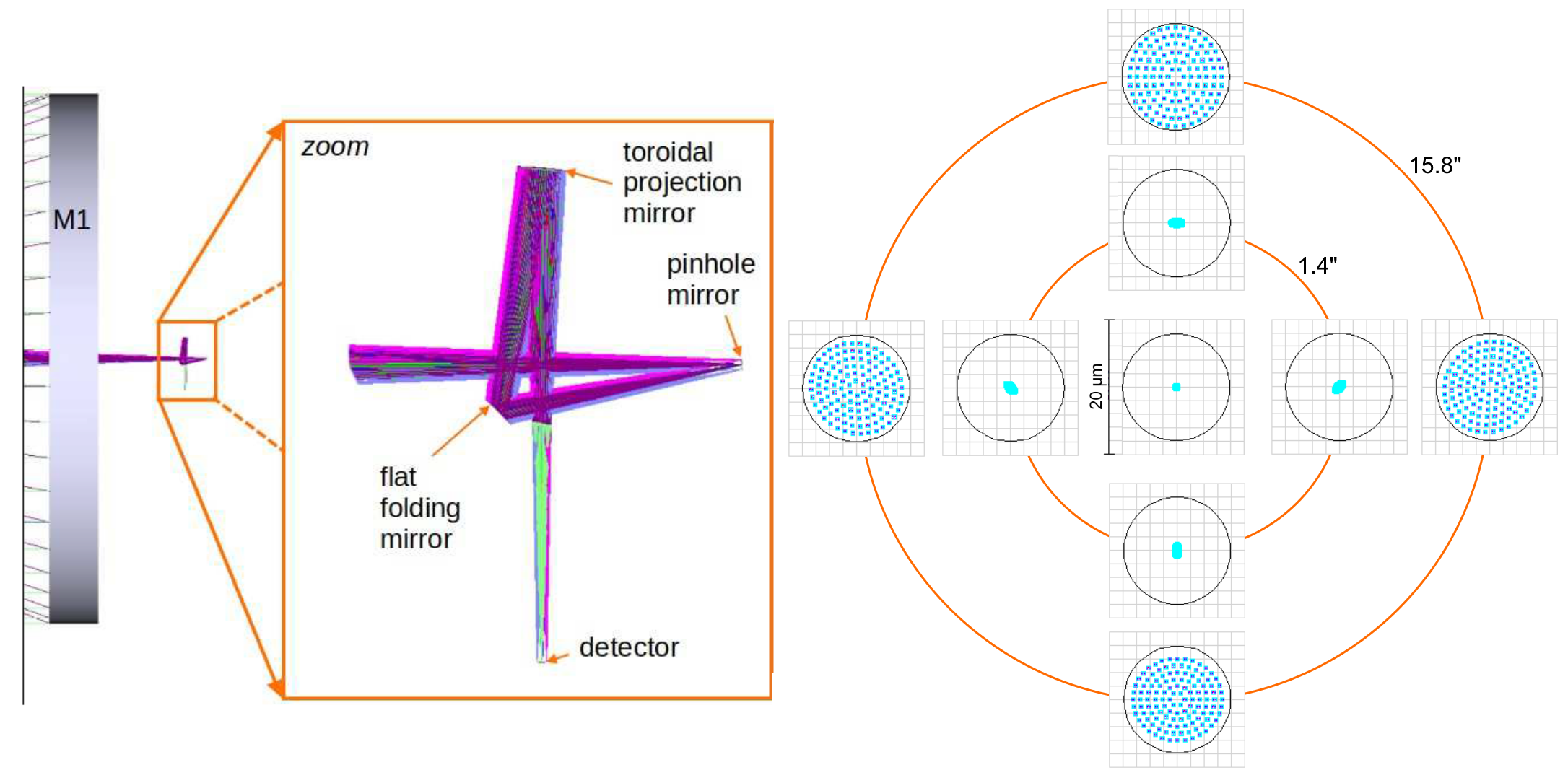}
   \end{tabular}
   \end{center}
   \caption[example] 
   { \label{fig:fgs1} 
First stage of the fine guiding system: left -- optical design, right -- spot diagrams.}
   \end{figure}

The FGS1 (see Fig.~\ref{fig:fgs1}) consists of a mirror of 2 mm in diameter, pierced with a $53~\mu m$ hole (field stop) in its centre, placed between the telescope and the polarimeter, facing the telescope, and inclined by $5.5^\circ$, which allows to send the tracking beam off-axis and keep the pinhole ellipticity negligible. The field of view of this mirror is $32"$, to include a $100\%$ margin on the platform pointing accuracy. A simple optical system is composed of two mirrors - one plane folding mirror with 7.4 mm diameter, and one toroidal mirror with 9.8 mm diameter, which projects the image with \textit{-1} magnification factor and astigmatism correction to the detector. The detector considered here is a $1024 \times 1024$ pixels CMOS with $2.5 ~\mu m$ pixel pitch. Using the information provided by the FGS1, the satellite can be more precisely pointed on the target, i.e. the target is recentered in the FGS1 field. A close-loop with the SVM allows to bring and maintain the target in the entrance hole with a \textit{200 mas} stability. 
As Fig.~\ref{fig:fgs1} shows, the projecting system provides a diffraction-limited quality for the entire field of view. So, the PSF is defined by the Airy disk size. Its FWHM is $6.7 ~\mu m$ and the pointing variation range is \textit{200 mas}, which corresponds to the PSF centre displacement of $ \pm 12.7 ~\mu m$ on the FGS1 detector. With the $2.5~ \mu m$ detector pixel size, the FWHM of the target thus covers $2.7 \times 2.7$ pixels and a \textit{200 mas} displacement corresponds to $\pm 5$ pixels. This provides enough information to precisely determine the centroid position of the PSF and correct pointing accordingly. 

\subsection{FGS stage 2}
\label{sec:fgs2}

\begin{figure} [ht]
   \begin{center}
   \begin{tabular}{c} 
   \includegraphics[width=13cm]{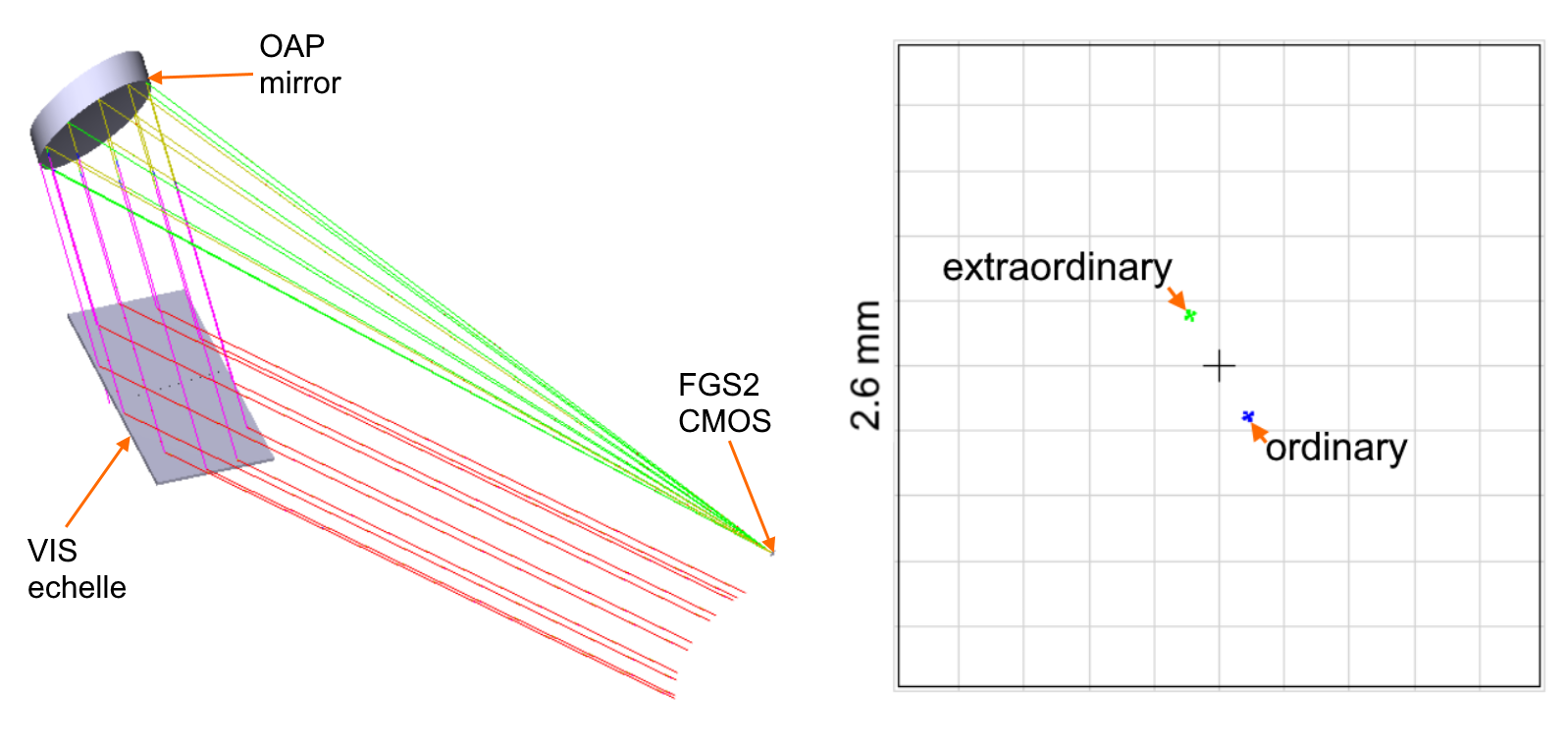}
   \end{tabular}
   \end{center}
   \caption[example] 
   { \label{fig:fgs2} 
Second stage of the fine guiding system: left -- optical design, right -- pinhole images at the detector.}
   \end{figure}

Once the FGS1 pointing sequence is complete, the target acquisition begins. The target light then enters the instrument and the FGS2. It consists of recording the $0^{th}$ diffraction order of the VIS echelle on a $1024 \times 1024$ pixel CMOS detector with $2.5 ~\mu m$ pixel pitch. The parallel beam is focused by an OAP mirror (\textit{60 mm} diameter, \textit{350 mm} focus, $50^\circ$ beam deviation), as it is shown in Fig.~\ref{fig:fgs2}. The $0^{th}$ order is anyway not used for science and it has the advantage of being recorded close to the focal plane, therefore the corrections from this FGS stage include some possible thermo-elastic deformations. The linear magnification factor between the pinhole and the FGS2 camera is \textit{-0.5}. The information recorded by the FGS2 is used to control a tip-tilt mirror placed between the polarimeter and the dichroic. The tip-tilt mirror applies the necessary correction to bring and maintain the $0^{th}$ order at a reference position at the centre of the detector. As a consequence the science orders of the VIS and UV spectrographs are also placed and maintained at the right position. A $\pm 200~ mas$ stability at the pinhole (provided by FGS1) produces a linear shift of the $0^{th}$ order spot of $\pm 12~ \mu m$ at the pinhole or $\pm 6 ~\mu m$ ($\pm 2.4$ pixels) on the FGS2 detector, $\pm 9.6 ~\mu m$ at the UV science detector, and $\pm 5.9~ \mu m$ at the VIS science detector. The correction of this shift requires a tip-tilt mirror rotation of $\pm 7 "$. Once this rotation has been executed, the remaining position error is below $1 ~\mu m$, as required, for the entire UV and Visible spectrum and the aberration blurring goes from less than $0.1 ~\mu m$ in the UV spectrum up to $0.7 ~\mu m$ at 888 nm. Since the light reaching the FGS2 detector passes through the polarimeter, the FGS2 detector records two $0^{th}$ order spots (for the two polarisations) separated by $464~ \mu m$ or 185.6 pixels (see Fig.~\ref{fig:fgs2}), therefore twice as much information is available to determine the position of the spots and the direction of the line between the centres of the two spots informs us about the spin of the satellite.

\section{CONCLUSIONS AND FUTURE WORK}
\label{sec:concl}  

The design of the \textit{Arago} mission payload optical system meets all the requirements derived from the science goals and the platform characteristics in terms of the spectral resolution, sensitivity, and pointing accuracy. It also meets key budget constraints, such as mass, volume, power, data with large
margins. Finally, it is mostly based on elements and units with high tehnological readiness level, which reduces the potential risks of the proposed mission.

In terms of the optical system development there are a few tasks to be considered during the next stages. In particular, a certain re-optimization of the spectrographs should be performed in order to compensate the residual dispersion of the polarimetric unit. The angles and distances, defining the positions of the dichroic, tip-tilt  mirror and two collimators may be slightly changed to provide a better packaging, manufacturability of the mechanical part, and facilitate the assembling. Depending on the choice of the technology to produce the diffractive components we may have to revise and slightly re-balance the throughput budget. Then, as any optical system, \textit{Arago} payload will require a careful modelling of the manufacturing and alignment errors and establishing of the end-to-end error budget. The interfaces with the mechanical structure and electronics units will require more attention as the corresponding workpackages become more elaborated. 

\section*{ACKNOWLEDGMENTS}       
 
The authors acknowledge the entire \textit{Arago} consortium for their contributions.  We also thank the PASO department of CNES for their help during the Phase 0 study, as well as Heather Bruce, Kenric Citadelle, and Ilian Ellafi at LESIA for their contribution. We are also thankful to ADS, Glyndwr Innovations, and SAFRAN/REOSC for their studies for \textit{Arago}, and to Horiba, Teledyne, and Zeiss for useful discussions and information.


\end{document}